\documentclass{article}
\usepackage{spconf,amsmath,graphicx}
\usepackage{booktabs}
\usepackage{caption}
\usepackage{hyperref}
\usepackage{diagbox} 
\usepackage[numbers, sort&compress]{natbib}
\usepackage{subfigure}
\usepackage{float}
\usepackage{balance}
\usepackage{makecell}
\usepackage{bibspacing}
\usepackage{lipsum}
\usepackage{bm}

\title{Unsupervised word-level prosody tagging for controllable speech synthesis}
\name{Yiwei Guo, Chenpeng Du, Kai Yu\sthanks{Corresponding author}}
\address{MoE Key Lab of Artificial Intelligence, AI Institute\\X-LANCE Lab, Department of Computer Science and Engineering\\Shanghai Jiao Tong University, Shanghai, China\\\texttt{\{cantabile\_kwok, duchenpeng, kai.yu\}@sjtu.edu.cn}}
\begin{document}
\ninept
\maketitle
\begin{abstract}

Although word-level prosody modeling in neural text-to-speech (TTS) has been investigated in recent research for diverse speech synthesis, 
it is still challenging to control speech synthesis manually without a specific reference.
This is largely due to lack of word-level prosody tags.
In this work, we propose a novel approach for unsupervised word-level prosody tagging with two stages, 
where we first group the words into different types with a decision tree according to their phonetic content and then cluster the prosodies using GMM within each type of words separately. 
This design is based on the assumption that the prosodies of different type of words, such as long or short words, should be tagged with different label sets. 
Furthermore, a TTS system with the derived word-level prosody tags is trained for controllable speech synthesis. 
Experiments on LJSpeech show that the TTS model trained with word-level prosody tags not only achieves better naturalness than a typical FastSpeech2 model, 
but also gains the ability to manipulate word-level prosody.

\end{abstract}
\vspace{-0.1in}
\section{Introduction}
\label{sec:intro}

\vspace{-0.1in}
Prosody modeling in neural speech synthesis has been extensively explored in recent research,
aiming for natural, diverse, and controllable synthesis.
The naturalness of synthetic speech is improved with prosody modeling taken into account \cite{natural2012, natural2014,natural2016}. 
Recently, more attention has been attracted by rich prosody modeling and control.

Explicit prosodic features, which have clear linguistic or phonological interpretation, are first investigated.
\cite{FineGrained,SemiSuper} both provide solutions to control specific acoustic aspects of phone-level speech.
\cite{FineGrained} introduces temporal structures in the embedding networks that can control pitch and amplitude either on speech side or text side. 
\cite{SemiSuper} proposes a generative model that controls affect and speaking rate with semi-supervised latent variables.
\cite{word-pitch} effectively controls word-level pitch accent by multiplying optional bias to pitch encoder's output. 
\cite{CtrlP, ControlExplicitFeature} presents F0, duration and energy control with variational auto-encoders (VAE). 
They disentangle these prosody features and provide more independent control.
\cite{UnsuperClustering, KoreaClustering} model these features with clustering, which is a purely data-driven method that have more interpretability.
In contrast to explicit representation, implicit prosody representation is more complete and richer when modelling prosody diversity, yet uninterpretable.
Prosody embeddings sampled from prior distribution with VAE are widely investigated in many linguistic levels.
\cite{utt-VAE} models the global characteristics for an utterance.
\cite{hsu2018hierarchical} improves the performance by incorporating GMM prior in VAE.
\cite{klimkov2019finegrained} enhances phone-level prosody latent representations by VAE in prosody transfer.
\cite{VQVAE} uses vector quantization and trains an autoregressive prior model to generate synthetic speech with better sound quality.
\cite{hie1,hie2,hie3} models prosody hierarchically, by conditioning phone and word-level latent variables on coarser ones.
These works incorporate more semantic information, thus improve the naturalness of synthetic speech to a great extent.
Recently, unsupervised prosody clustering with mixture density network is also proposed in \cite{cpd-trans,cpd-interspeech}, enabling richer prosody diversity.

However, all the prior works control the prosodies manually by providing a reference speech or specifying the values of explicit prosodic features, such as pitch, which is hard to be practically applied. For example, it is expensive to collect reference speech with the prosodies that one needs. 
Also, hand-written values of explicit features may not correspond to a natural speech, and these explicit features do not represent the entire prosody space. 
As for implicit prosody representations, there are few known methods that can control prosody in inference stage. This is mainly because of the continuous prosody distributions they use. Therefore, few of the existing works achieve good and interpretable controllability with diverse prosody in natural speech.
In this work, we propose an unsupervised word-level prosody tagging system that can be directly used for prosody control. 
We extract prosody embeddings from the mel-spectrogram of reference speech. 
Then, we obtain the word-level prosody tags in two stages.
First, we construct a decision tree that recursively clusters all the words into different text-dependent sets, with a set of questions regarding their phonetic contents.
Then, for each text-dependent leaf node, we cluster the prosody embeddings using Gaussian mixture models.
The obtained prosody tags represent word-level prosody types and are further embedded to train a TTS system with a prosody tag predictor.
The prosody tag predictor is capable of controlling the prosody of synthetic speech by manually specifying the prosody tag of any word.

Our approach has several advantages besides the improved naturalness and controllability.
First, the prosody tags are obtained in an unsupervised manner,
without the need for expensive manual annotations like emotional labels.
Second, the decision tree design makes it easy and robust to generalize to unseen words in inference, in terms of identifying a word into its phonetic cluster.
Furthermore, as most of the questions in decision tree are language-agnostic, this design can be easily extended to different languages. 
By selecting the questions, the tree can also be used for multiple tasks.
\begin{figure*}[t!]
\centering
    \includegraphics[scale=0.19]{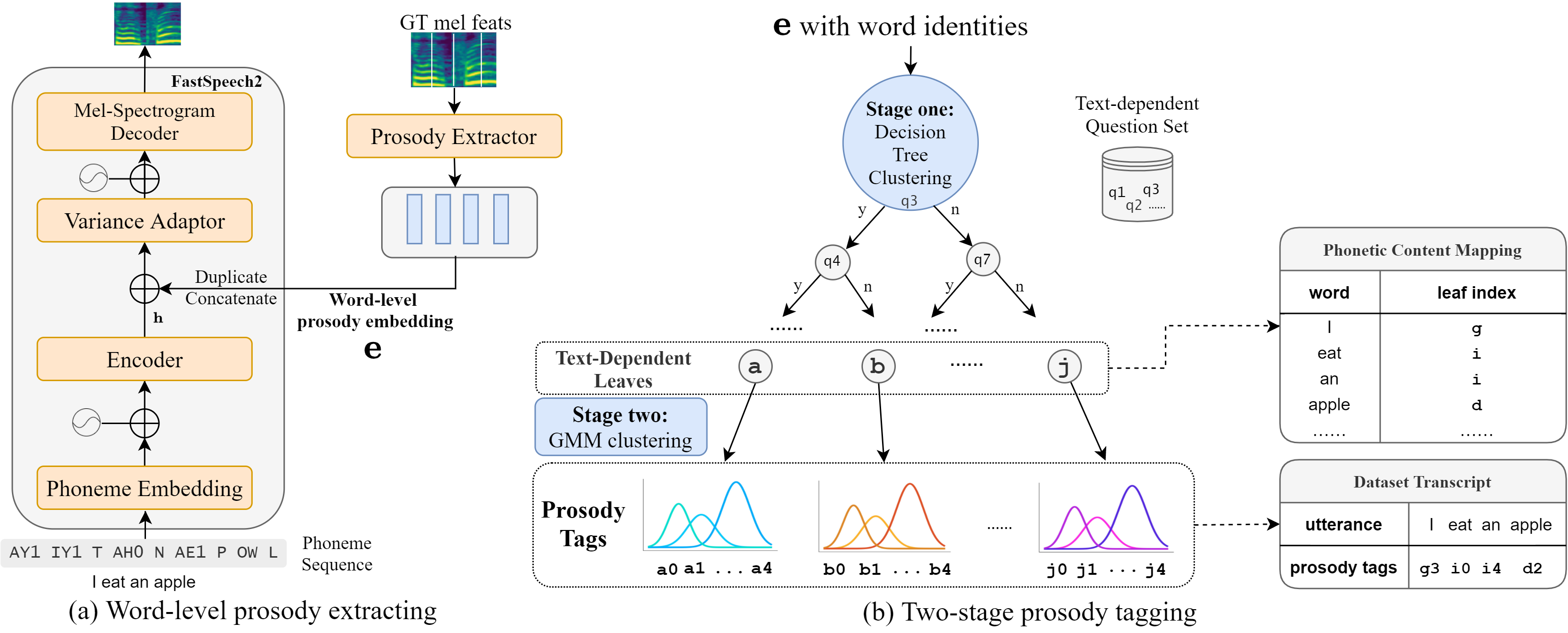}
    \vspace{-5pt}
    \caption{Prosody extracting and tagging system architecture}
    \label{fig:extractor_and_tree}
\end{figure*}

The rest of the paper is organized as follows. Section \ref{sec:model} illustrates the overall system. 
Experiments and results analysis are given in Section \ref{sec:exp}, and Section \ref{sec:conclu} draws a conclusion.

\section{word-level Prosody tagging and control}
\label{sec:model}

Our system is built in three steps: word-level prosody embedding extracting, two-stage word-level prosody tagging, and TTS training with the prosody tags. Note that the TTS models in our system are based on FastSpeech2 \cite{Fastspeech2}.

\vspace{-0.1in}
\subsection{Word-level prosody extracting}

In order to obtain word-level prosody embeddings, we first build a typical FastSpeech2-based TTS model together with a prosody extractor following \cite{cpd-interspeech}. As is shown in Fig.\ref{fig:extractor_and_tree}(a), the prosody extractor generates a hidden vector (named as prosody embedding $\mathbf e$) for each word from the corresponding mel-spectrogram segment. The generated prosody embeddings are then aligned with the phoneme sequence and concatenated to the encoder output. Accordingly, the extractor is optimized to extract useful information for better reconstructing the output speech, including both prosody information and phonetic contents of the words.

\begin{figure*}[h]
    \centering
    \includegraphics[scale=0.14]{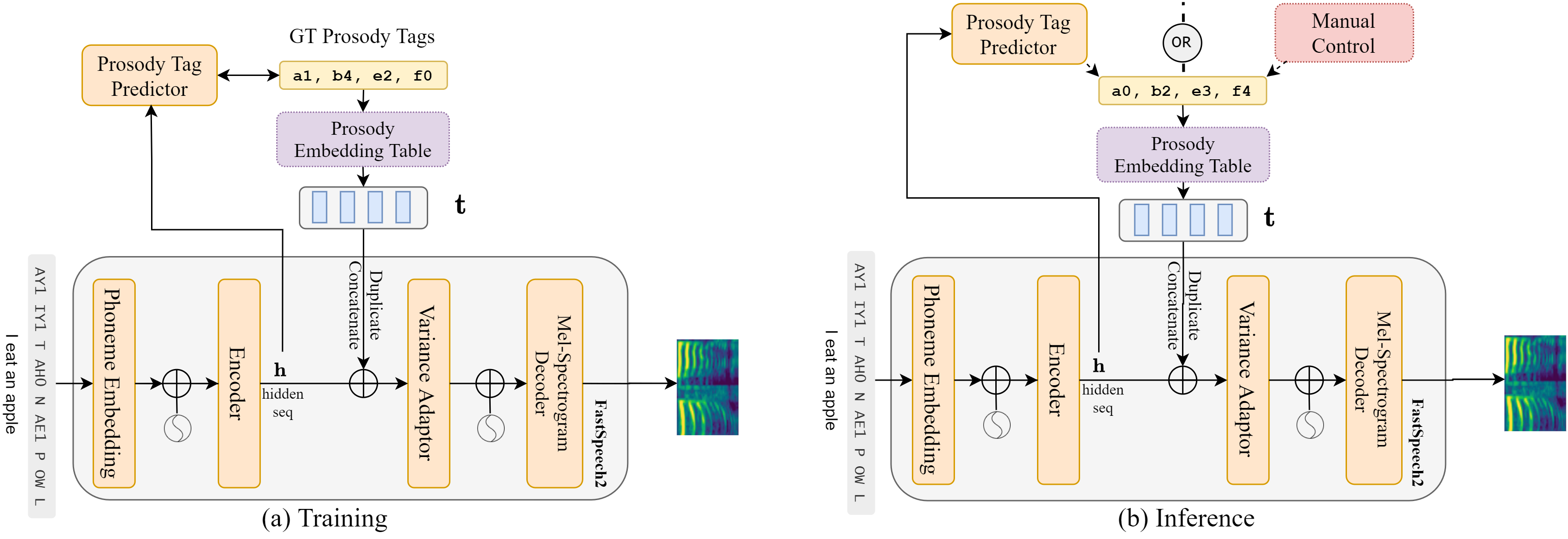}
    \caption{Prosody control model architecture in training and inference stage}
    \label{fig:predictor_train_inference}
\end{figure*}


\vspace{-0.1in}
\subsection{Prosody tagging with two stages}



It is an intuitive idea that words with greatly different phonetic contents, such as the long word 'congratulation' and the short word 'cat', are uttered in a completely different ways and consequently should not be tagged with the same set of prosody tags. Therefore, in this work, we design a two-stage prosody tagging strategy, where we first group the words into different types with a decision tree according to their phonetic contents and then cluster the prosodies using GMM within each type of words separately.

\vspace{-.1in}
\subsubsection{Stage one: decision tree clustering}

Following the HMM state-tying in ASR \cite{tree}, we construct a binary decision tree for word clustering with a set of questions $Q$ on its phonetic contents, where all the words in the root are clustered into $l$ leaves. 
These questions are designed based on our expert knowledge, such as "Whether the phonemes of the word are more than 4 or not?"
and "Whether the word ends with a closed syllable?". 
We reference the phonetic questions in HTS\cite{HTS}, which is a direct product of \cite{tree}.

Each node in the decision tree contains a set of words whose prosody embeddings can be modeled with a Gaussian distribution and the log-likelihood can be formulated as 
\begin{equation}
    LL^{(i)} =\sum_{\mathbf e \in \mathcal{E}^{(i)}}\mathcal \log \, {\cal N}\left(\mathbf e \mid{\bm\mu}^{(i)},{\bm\Sigma}^{(i)}
    \right) 
\end{equation}
where $i$ is the node index and $\mathcal{E}^{(i)}$ is the set of all prosody embeddings corresponding to the words in the node $i$.
Each non-leaf node $i$ is related to a question $q$ that partitions the words in the node into its left or right child, leading to an increase in log-likelihood of the prosody embeddings
\begin{equation}
    \Delta_q LL^{(i)}=LL^{(i\text{'s left child under }q)}+LL^{(i\text{'s right child under }q)}-LL^{(i)}.
\end{equation}


The initial tree contains only a root node, which is also a leaf node. Then we recursively perform the following step: find the question that maximizes the increase in log-likelihood for all the leaf nodes, and select a leaf node $j$ whose increase is the maximum over all the leaf nodes, which is
\begin{equation}
    j = \arg\max_{i \in \text{leaf nodes }} \left(\max_{q\in Q} \Delta_q LL^{(i)}\right)\label{split_criterion},
\end{equation}
 and split the selected node with the corresponding question. This process continues until the increase in log-likelihood is smaller than a threshold. Consequently, the topology of the decision tree is obtained. In this work, the number of leaves $l$ is 10 as shown in Fig. \ref{fig:extractor_and_tree}(b), whose indices are denoted as letters from $\texttt a$ to \texttt j.

\vspace{-.18in}
\subsubsection{Stage two: Gaussian mixture clustering}

The word-level prosody embeddings $\mathbf{e}$ extracted by neural networks contain both prosody information and phonetic content of the words. However, the decision tree clusters the words into $l$ leaves according to the questions only on their phonetic contents, so we assume that the prosody embeddings of the words in a leaf node differ only in prosodies and are similar in phonetic contents. Therefore, clustering within a leaf node is dominated by the prosodies instead of phonetic contents.

We perform GMM-based clustering for the prosody embeddings within each leaf node $i$ separately, which is
\begin{equation}
    \mathbf e^{(i)}\sim \sum_{k=1}^m \omega_k^{(i)}\mathcal N\left(\mathbf e^{(i)}|{\bm\mu}_{k}^{(i)}, {\bm\Sigma}_k^{(i)}\right)
\end{equation}
where $k$ is the Gaussian component index and $m$ is the number of components.
The prosody of each word is tagged with the index of the Gaussian component that maximizes the posterior probability of its prosody embedding $\mathbf{e}$
\begin{equation}
    t=\arg\max_{k} \left\{\log \mathcal N\left(\mathbf e\mid{\bm\mu}^{(i)}_k, {\bm\Sigma}^{(i)}_k\right)+\log \omega_k^{(i)}\right\}.
\end{equation}
In this work, $m$ is set to 5, so the Gaussian component ids range from 0 to 4. Accordingly, all the words in the training set are labelled with the $m\times l = 5 \times 10 = 50$ prosody tags, which is the combination of $10$ leaf ids and $5$ Gaussian component ids. As shown in Fig. \ref{fig:extractor_and_tree}(b), the prosody tags are from $\texttt{a0}$ to $\texttt{j4}$.

Note that our prosody extracting and tagging system is fully unsupervised in which only audio information is utilized. Also, the tagging system is driven by both data and knowledge.


\vspace{-.14in}
\subsection{Prosody control with prosody tags}


Finally, we train a TTS model with the derived word-level prosody tags as shown in Fig.\ref{fig:predictor_train_inference}. In the training stage, the TTS model is guided by prosody embeddings retrieved from a trainable embedding table given the ground-truth prosody tags. In the inference stage, the prosody tags can be either predicted from input text by a prosody predictor or be manually specified. 


The prosody predictor in this work is similar to \cite{cpd-interspeech}. It predicts the prosody tag for each word given its corresponding phoneme hidden states, i.e. the encoder output sequence $\mathbf{h}$. The prosody predictor contains a bi-GRU that transforms the phoneme hidden states to a vector for each word, two convolutional blocks and a softmax layer. The convolutional blocks here consist of a 1D convolutional layer followed by a ReLU activation layer, layer normalization, and a dropout layer. The predictor is optimized by the cross-entropy loss $\mathcal L_{\texttt{PP}}$ with the ground-truth prosody tags.
Hence, the overall loss for the model training is defined as 

\begin{equation}
    \mathcal L = \alpha \mathcal L_{\texttt{PP}} + \mathcal L_{\texttt{FastSpeech2}}\label{loss},
\end{equation}
where $\mathcal L_{\texttt{FastSpeech2}}$ is the loss of FastSpeech2\cite{Fastspeech2} and $\alpha$ is the relative weight between the two terms.


\section{experiments and results}
\label{sec:exp}
\vspace{-0.05in}
\subsection{Experimental setup}
We use LJSpeech \cite{ljspeech}, a single-speaker dataset containing about 24 hours of recordings for our experiments.
242 utterances are left out as a test set. 
All utterances are down-sampled to 16kHz. 
We use 800-point window length, 200-point hop size, 1024 FFT points, and 320 mel-bins for feature extraction.
The phoneme alignment is obtained from an HMM-GMM ASR model trained on Librispeech \cite{librispeech}. 
The vocoder used in this work is MelGAN \cite{melgan}.
The coefficient $\alpha$ in Eq.\eqref{loss} is set to $1.0$.
The prosody embedding $\mathbf e$ is 128 dimensional. 

\vspace{-0.1in}
\subsection{The performance of decision tree in prosody tagging}
\label{exp:tree}


\vspace{-0.1in}
\begin{figure}[htbp!]
    \centering
    \includegraphics[scale=.33]{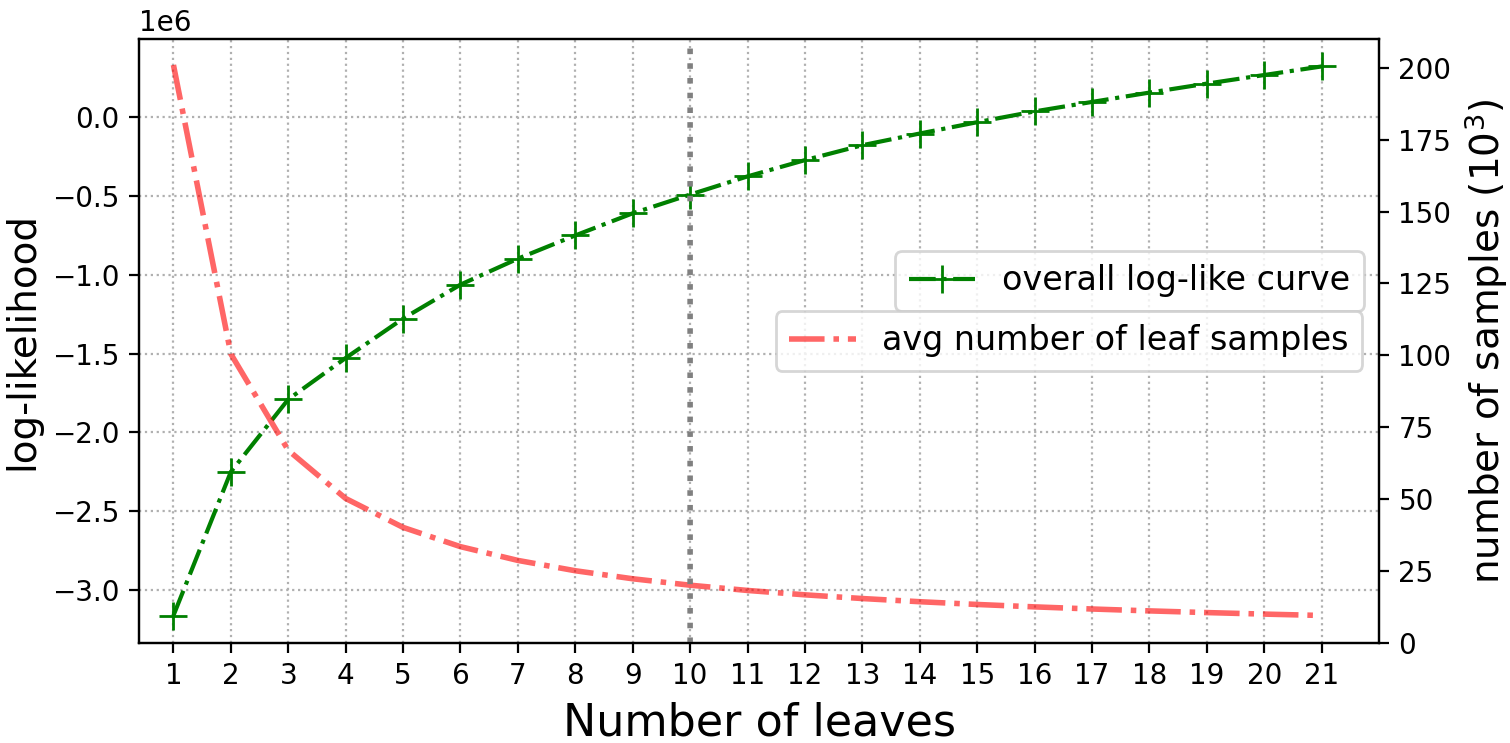}
    \caption{Curve of overall log-likelihood of leaves and average number of leaf samples}
    \label{fig:tree}
\end{figure}
\vspace{-0.1in}

We demonstrate the curve of the average number of prosody embeddings in each leaf node and the overall log-likelihood of prosody embeddings over all leaf nodes $\sum_{i \in \text{leaf nodes}}LL^{(i)}$ in Fig.\ref{fig:tree} when the tree grows. With the increase of the number of leaf nodes, the average number of prosody embeddings in each leaf node decreases whilst the overall log-likelihood of prosody embeddings increases. We stop the growth of the tree when the number of leaves reaches 10, in consideration of both the performance and the complexity.

\vspace{-0.12in}
\subsection{Naturalness of predicted prosodies}
\begin{figure}
    \centering
    \includegraphics[scale=.33]{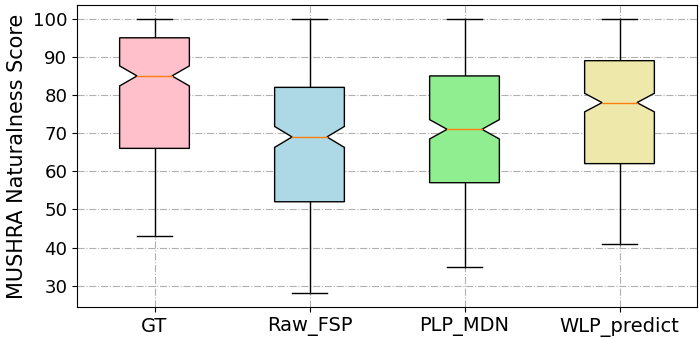}
    \caption{Subjective evaluation of naturalness}
    \label{fig:naturalness}
\end{figure}

The TTS model with a prosody predictor is trained with the derived word-level prosody tags. In the inference stage, the word-level prosodies can be either predicted from the input text by the prosody predictor or be manually specified. In this section, we synthesize the test set whose prosodies are predicted and sampled. Then we evaluate the naturalness with a MUSHRA test in which 30 listeners are asked to rate each utterance in a range from 0 to 100. We compare our model with two baselines: the typical FastSpeech2 model \cite{Fastspeech2} Raw\_FSP and a TTS model in which phone-level prosodies are modeled with a mixture density network \cite{cpd-interspeech} PLP\_MDN. Also, the ground-truth mel-spectrograms of the recordings are reconstructed by the vocoder and then provided as GT in the listening test. The results are reported in Fig.\ref{fig:naturalness}. 
It can be observed that
our proposed word-level prosody prediction system with predicted prosody tags (WLP\_predict) outperforms both other models in terms of naturalness, due to our word-level prosody modelling, 
although it is still slightly worse than GT.

\vspace{-0.14in}

\subsection{Prosody controllability}

In order to evaluate the word-level prosody controllability of our TTS model, we first label the ground-truth word prosodies for the test set with the proposed prosody tagging system. Then we synthesize the test set 5 times where the prosody tags of the words in leaf $\texttt d$ are manually specified as $\texttt{d0}$ to $\texttt{d4}$ respectively while the prosody tags of other words are predicted and sampled. 
\footnote{The audio examples are available at  \url{https://cantabile-kwok.github.io/word-level-prosody-tagging-control/}}

\begin{figure}[H]
    \centering
    \includegraphics[scale=.38]{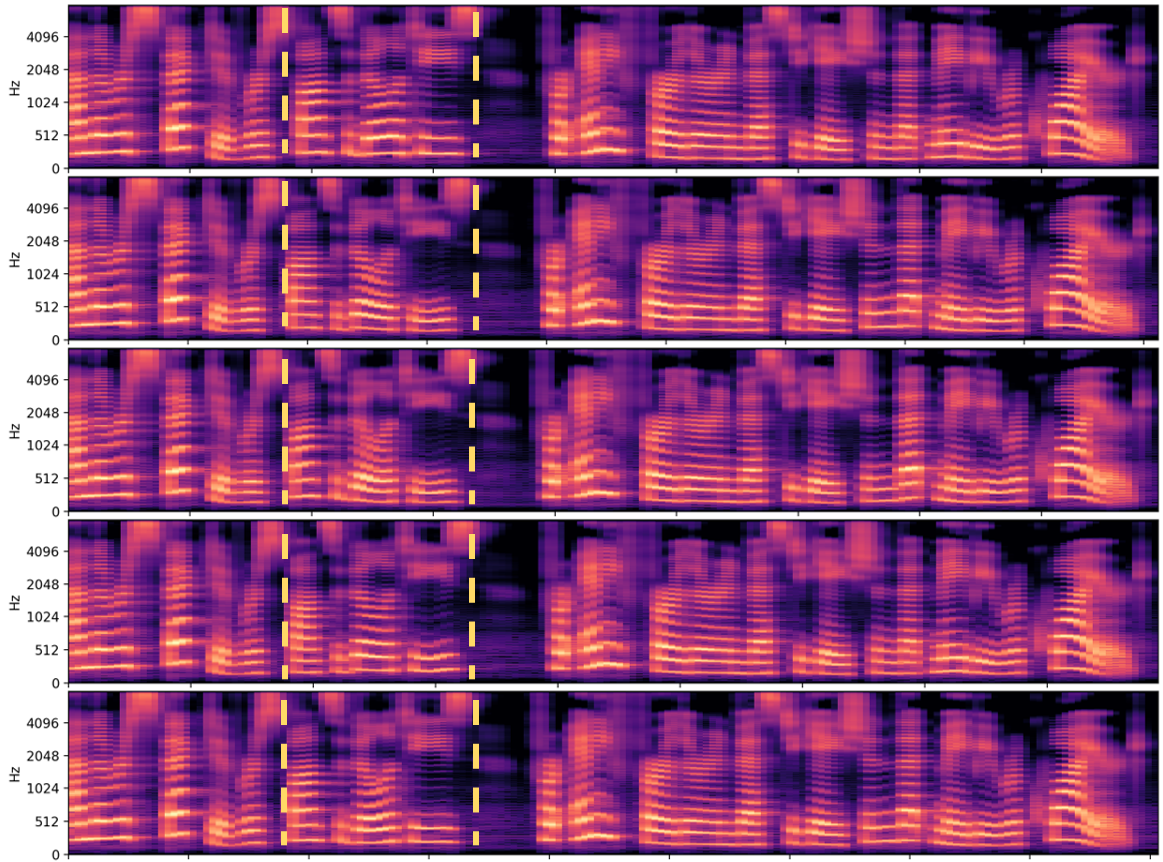}
    \vspace{-5pt}
    \caption{
    An example of synthetic speech with manually specified prosodies. The word between the yellow dash lines is ``responsibilities" whose prosody tags are specified as $\texttt{d0}$ to $\texttt{d4}$ respectively.
    }
    \label{fig:mel}
\end{figure}

Fig. \ref{fig:mel} shows an example in which the word ``responsibilities'' between the  yellow dash lines are manually controlled with $\texttt{d0}$ to $\texttt{d4}$ respectively. It can be observed that all the 5 prosodies of the word are different, showing the controllability of the prosody tags.

In addition, we need to confirm that same prosody tags lead to similar prosodies. Therefore, we evaluate the prosody similarity between the recordings and the synthetic speech with different specified prosody tags for all the words in the leaf $\texttt d$ in the test set. Theoretically, when the specified prosody tag is equal to the ground-truth prosody tag, the word prosody in the synthetic speech should be most similar to the recordings.

We perform the evaluation of prosody similarity in objective and subjective ways respectively. We first compute the average Mel cepstral distortion (MCD) over all the words with ground-truth prosody tag $\texttt{d}t$ where $t$ ranges from 0 to 4 between the recordings and the synthetic speech with a certain specified prosody tag. The results are reported in Table \ref{table:mcd}. As expected, we can find that all the diagonal values are the lowest among the values on their columns, showing that same prosody tags lead to similar prosodies in synthetic speech.

\begin{table}[t]
\begin{tabular}{c|c|c|c|c|c}
\toprule[1pt]
\diagbox{\makecell{\footnotesize{Ctrl Tag}}}{\footnotesize{GT Tag}} & 0                                  & 1                                  & 2              & 3              & 4              \\ \hline
0                        & \textbf{5.389} & 5.434          & 5.371          & 5.490          & 5.420          \\
1                        & 5.410          & \textbf{5.348} & 5.379          & 5.796          & 5.420          \\
2                        & 5.612                              & 5.670                              & \textbf{5.356} & 5.548          & 5.517          \\
3                        & 5.828                              & 6.023                              & 5.578          & \textbf{5.442} & 5.714          \\
4                        & 5.507                              & 5.507                              & 5.362          & 5.562          & \textbf{5.309} \\ \bottomrule[1pt]
\end{tabular}
\caption{Mel cepstral distortion between the recordings and the synthetic speech with different specified prosody tags for all the words in the leaf $\texttt d$ in the test set.}
\label{table:mcd}
\end{table}

\begin{figure}[H]
    \centering
    \includegraphics[scale=.28]{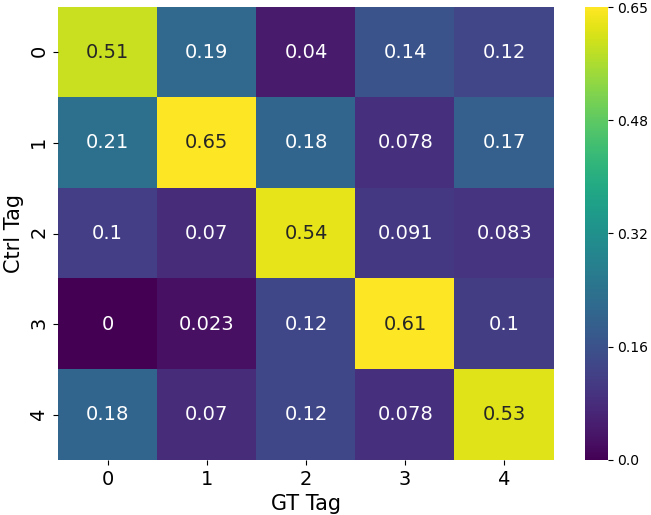}
    \caption{Subjective evaluation of controllability}
    \label{fig:matrix}
\end{figure}

Also, we evaluate the prosody similarity with a subjective listening test where 30 listeners are provided with the recording and 5 synthetic speech with different prosody tags for each group and are asked to select the synthetic speech whose prosody of the corresponding word is the most similar to the recording.
The proportions of the selections are depicted as a confusion matrix in Fig. \ref{fig:matrix}. Similar to the results of objective evaluation, the proportion of the synthetic speech with the same prosody tags to the ground-truth ones, i.e. the diagonal values, achieves the highest among their columns, which further confirms the controllability of prosody tags.

\vspace{-0.15in}



\vspace{-0.16in}
\section{conclusion}
\label{sec:conclu}

In this work, we propose a novel approach for unsupervised word-level prosody tagging with two stages, 
where we first group the words into different types with a decision tree according to their phonetic content and then cluster the prosodies using GMM within each type of words separately. 
Furthermore, a TTS system with the derived word-level prosody tags is trained for controllable speech synthesis, where the prosody can be either predicted from input text or manually specified.
Experiments on LJSpeech show that our model achieves better naturalness than a typical FastSpeech2 model with the predicted prosodies. In addition, the objective and subjective evaluations for prosody controllability show that the prosodies can be efficiently controlled by specifying the word-level prosody tags.

\vspace{-0.1in}
\section{ACknowledgements}
This study was supported by State Key Laboratory of Media Convergence Production Technology and Systems Project (No. SKLMCPTS2020003) and Shanghai Municipal Science and Technology Major Project (2021SHZDZX0102).


\vfill\pagebreak

\bibliographystyle{IEEEbib}
\bibliography{strings,refs}


\end{document}